\documentclass[a4paper,fleqn,usenatbib,useAMS]{mnras}
\usepackage{graphicx}
\usepackage[skip=2pt,font=footnotesize]{caption}
\usepackage [autostyle, english = american]{csquotes}
\usepackage{natbib}
\usepackage{lipsum}
\usepackage{mathtools}
\usepackage{cuted}
\usepackage{capt-of}
\usepackage{tabu}
\usepackage{float}
\usepackage{hyperref}
\usepackage{url}
\usepackage{color}
\usepackage[T1]{fontenc}
\usepackage{ae,aecompl}

\title{Prospects of Probing Dark Energy with eLISA: Standard versus Null Diagnostics}

\author[Baral et al.]{Pratyusava Baral,$^{1}$\thanks{E-mail: \href{baralpratyusava@gmail.com}{baralpratyusava@gmail.com}}
Soumendra Kishore Roy$^{1}$\thanks{E-mail: \href{soumendrakishoreroy09@gmail.com}{soumendrakishoreroy09@gmail.com}} and Supratik Pal$^{2}$\thanks{E-mail: \href{supratik@isical.ac.in}{supratik@isical.ac.in}}\\
$^{1}$Department of Physics, Presidency University, 86/1 College street, Kolkata 700073, India.\\
$^{2}$Physics \& Applied Mathematics Unit, Indian Statistical Institute, 203 B. T. Road, Kolkata 700108, India}

\hypersetup{
        colorlinks=true,
        linktoc=all,
        linkcolor=blue,
        citecolor=blue,
        filecolor=black
}
\begin{document}

\maketitle

\begin{abstract}
Gravitational waves from supermassive black hole binary mergers along with an electromagnetic counterpart have the potential to shed `light' on the nature of dark energy in the intermediate redshift regime. Accurate measurement of dark energy parameters at intermediate redshift is extremely essential to improve our understanding of dark energy, and to possibly resolve a couple of tensions involving cosmological parameters. We present a Fisher matrix forecast analysis in the context of eLISA to predict the errors for three different cases: the non-interacting dark energy with constant and evolving equation of state (EoS), and the interacting dark sectors with a generalized parametrization. In all three cases, we perform the analysis for two separate formalisms, namely, the standard EoS formalism and the \textit{Om} parametrization which is a model-independent null diagnostic for a wide range of fiducial values in both phantom and non-phantom regions, to make a comparative analysis between the prospects of these two diagnostics in eLISA.  Our analysis reveals that it is wiser and more effective to probe the null diagnostic instead of the standard EoS parameters for any possible signature of dark energy at intermediate redshift measurements like eLISA.
\end{abstract}

\begin{keywords}
Gravitational Wave Cosmology: Dark Energy Equation of State -- Om Parameter -- Supermassive Black Hole Binary Merger
\end{keywords}

\section{Introduction}
The Universe at large as we know today is filled with a component exerting negative pressure (called Dark Energy), which pushes everything further and further away from us. Although this late-time acceleration \citep{Perlmutter_1997, Perlmutter_1999, Riess_1998, Astier_2006} has been known for more than two decades, a completely satisfactory theoretical perspective that fits with all observations is yet to be achieved. Within General Relativity, the widely used model today  is $\Lambda$CDM where the dark matter is expected to be non-interacting and ``cold" with an equation of state (EoS henceforth) $w_{\rm dm}=0$ and dark energy is identified with the cosmological constant $\Lambda$ having the EoS, $w_{\rm de}=-1$ \citep{Sahni_2008, Weinberg_1989, Bousso_2007}.  However, this has several caveats. Along with the well-known theoretical issues such as the fine-tuning problem and the cosmic coincidence problem, some unavoidable observational issues have emerged of late. The direct measurement of the Hubble constant for local galaxies by \citet{Riess_2016} gave $H_0 = (73.24 \pm 1.74)$ km/s/Mpc which is in tension at $\sim 3.4\sigma$ with the result derived from CMB ($z \sim 1100$). CMB measurements give $H_0 = (67.4 \pm 0.5)$ km/s/Mpc \citep{collaboration2018planck}, under the assumption of $\Lambda$CDM model and three flavors of neutrino. This tension was first pointed out after the Planck 2013 data release \citep{PhysRevD.91.083005, Novosyadlyj_2014}. With more data coming from Planck as well as from direct measurements of $H_0$, the tension has proved to be larger than ever. This result has been supported by weak lensing time delay experiments \citep{wong2019h0licow} that further raise the tension to $5.3 \sigma$ for the joint analysis with the time delay cosmography and the distance ladder results.  Disagreements have also been found between local measurements and cosmological measurements of the root mean square density fluctuation, $S_8 = \sigma_8 (\Omega_{0m}/3)^{0.5}$, and the present abundance of the dark matter, $\Omega_{0m}$. Sunyaev-Zeldovich cluster counts ($S_8 = 0.78 \pm 0.01$) \citep{2014}, DES ($S_8 = 0.783^{+0.021}_{-0.025}$) \citep{Abbott_2018} and KiDS-450 weak lensing surveys ($S_8 = 0.745 \pm 0.039$) \citep{Hildebrandt_2016} consistently report a smaller value ($\sim 2 \sigma$) of matter fluctuation than that reported in Planck ($S_8 = 0.811 \pm 0.006$) \citep{collaboration2018planck}. Regarding $\Omega_{0m}$, BOSS measurement of Lyman-$\alpha$ forest \citep{PhysRevD.92.123516} and DES \citep{Abbott_2018} favor a smaller $\Omega_{ 0m}$ than the Planck Collaboration \citep{collaboration2018planck}. The measurements from distant quasars also show the departure from $\Lambda$CDM at high redshifts with $\sim 4\sigma$ confidence \citep{risaliti2018cosmological} and agrees with $\Lambda$CDM at low redshifts ($z<1.4$). Altogether these show the tension between low and high redshift measurements are generic to the $\Lambda$CDM model \citep{Bhattacharyya_2019} and it is very difficult to blame the systematics for this discrepancy \citep{10.1093/mnras/stu278, Addison_2016, 2017, Aylor_2019, PhysRevD.102.023518}. 

To unravel this concern, theorists have considered different types of dark matter and dark energy fluid \citep{COPELAND_2006, Zlatev_1999, Armendariz_Picon_2000, PhysRevD.79.043502, Di_Valentino_2016, Di_Valentino_2017, Di_Valentino_2017_1, Poulin_2019} both with and without interaction between them \citep{PhysRevD.101.063502, valentino2019interacting, Costa_2017, van_de_Bruck_2019}. Proposals with modifications of  GR,  that generically go by the name modified gravity, are also around.  \citep{capozziello2003quintessence, Carroll_2004, Sotiriou_2010, Nojiri_2004}. To confront this wide spectrum of models with observations,  a large class of dynamical dark energy models and several modified gravity theories are represented by some generic parametrizations (e.g. $w_0$CDM for constant EoS and CPLCDM that can account for redshift evolution of dark energy EoS, if any). The local and CMB measurements have put stringent constraints on these parameters \citep{Cai_2010, PhysRevD.88.063501, PhysRevD.92.123503, refId0}. However, even with these tight
constraints, several models are still allowed, all of them resembling $\Lambda$CDM with proximity. As it turns out, we cannot have significant improvement further about our knowledge of dark energy than what we have so far, using present datasets.
Herein lies the importance of probing intermediate redshifts that have the potential to reflect either redshift evolution of dark energy EoS, if any, or information about dark energy perturbations in form of a non-trivial sound speed or cosmic shear.  Two very crucial upcoming missions using electromagnetic astronomy that target to probe these features, among others, at intermediate redshifts are Square Kilometre Array (SKA) \citep{bacon_battye} and Thirty Meter Telescope (TMT) \citep{Skidmore_2015}.

Gravitational waves (GW) from standard sirens \citep{Klein_2016} like supermassive black hole binary mergers is another unique way of looking at middle redshifts \citep{Tamanini_2016}. The main advantage of using gravitational waves is that it can break the degeneracy between GR and modified gravity pretty well \citep{Belgacem_2018}. It can also provide strong constraints on the nature of the dark sector \citep{Tamanini_2016}. Additionally, it has the potential to resolve the said tension with the Hubble parameter.  
In the upcoming days, the space-based gravitational wave observatory  Laser Interferometer Space Antenna (LISA), primarily led by the European Space Agency (ESA) \citep{eLISA}, is expected to detect the supermassive black hole mergers at relatively high redshifts. This three-arm interferometer will orbit the Sun at a radius of $1AU$ and is capable of measuring mid-frequency gravitational waves around mHz regime.  The cosmography using LISA is based on extracting the gravitational wave luminosity distance from the waveform of the supermassive black hole binary, and its redshift is detected from electromagnetic counterparts \citep{Klein_2016}. The fact that gravitational wave can be of extreme use in cosmology is known since the late 1980s \citep{Schutz, Holz_2005, Cutler_2009}. The recent detection of GW170817 by the LIGO-VIRGO collaboration along with a coincident detection of its electromagnetic counterpart has finally opened the field of gravitational wave cosmology \citep{Abbott_2017, Coulter_2017, Goldstein_2017, Savchenko_2017}. Using this an independent measurement of $H_0=70.0 ^{+12.0}_{-8.0}$ km /s /Mpc \citep{2017_1} has been made which is consistent with both local and CMB measurements. In the post LIGO scenario, it has been established that eLISA (today's version of LISA) can probe the acceleration of the Universe \citep{Tamanini_2016} provided there exists an electromagnetic counterpart to detect the redshift. Since eLISA has not flown yet, errors have been forecasted for various parameters in a different class of dark energy models, such as late \citep{Tamanini_2016}, interacting, and early dark energy \citep{Caprini_2016} with the $\Lambda$CDM fiducial values and with standard EoS parametrizations. 

On the other hand, to examine the departure from the $\Lambda$CDM model, Sahni et al. \citep{OmSahni} developed a model-independent null diagnostic, called the {\it Om} parameter, which can probe dark energy directly from observational data without any reference to $\Omega_{0m}$. Unlike the standard parametrization where an erroneous choice of $\Omega_{0m}$ can make a fiducial $\Lambda$CDM universe, phantom, or quintessence, {\it Om} is naturally immune to the present matter density. Moreover, {\it Om} involves the first derivative of the luminosity distance which causes much less numerical error than the direct determination of the EoS parameter as this is a function of the second derivative of luminosity distance. {\it Om} has even been modified a little bit so that the Baryon Acoustic Oscillation (BAO) data can also be analyzed using this parameter \citep{Shafieloo_2012}. However, based on the present dataset, this parameter has shown no significant deviation from $\Lambda$CDM which is because only the local Universe at relatively low redshifts has been probed using this parameter to date.

In the present article, our primary intention is to investigate for prospects of probing dark energy using gravitational wave standard sirens in eLISA and to make a comparison between  standard EoS parametrization and {\it Om} diagnostics. 
As eLISA is going to detect supermassive black hole mergers at intermediate redshifts, we presume that {\it Om} might be a good parametrization to probe the nature of dark energy. To accomplish this, we do a Fisher matrix forecast analysis for standard EoS parametrization and {\it Om} diagnostics, which would help us to make a comparison between errors for the two distinct formalisms in the light of eLISA.

Our analysis is based on forecast of errors for standard vis-a-vis {\it Om} parametrizations in two widely used non-interacting dark energy parametrizations, namely, the $w_0$CDM with a constant EoS for dark energy and CPLCDM (where dark energy EoS is parametrized as  $w=w_0 + w_ a\frac{z}{1+z}$, so a two-parameter description), as well as for a generalized setup for interacting dark sectors that can have an {\it effective} EoS for dark matter as well, along with the dark energy EoS. As already demonstrated, the last one can in principle boil down to different (non)interacting models including warm dark matter, for a suitable choice of EoS parameters.
For our forecast, we make use of a wide range of fiducial values chosen from the constraints coming out of existing data for Planck 2015 + R16 and Planck 2015 + BSH \citep{Bhattacharyya_2019}. 
For each class of models, we analyze in terms of both EoS and {\it Om} parametrization and then compare between the errors. 
Our analysis reveals that  the {\it Om} parameter is indeed a better choice in terms of the error to constrain the dark sectors using standard sirens in eLISA. 
We also find {\it Om} can constrain phantom equally well as quintessence at least for a range of fiducial values whereas standard parametrization always favors quintessence models. {\it Om} also offers much less error on $w_a$ for the CPLCDM parametrization than the same in the standard technique. Further, along with forecasts on dark energy, we also give possible constraints on the error of deviation of dark matter from its `coldness' that may arise either from warm dark matter or from the interaction between the dark sectors via an {\it effective} EoS. Throughout the paper, we are assuming the Universe to be spatially flat ($\Omega_{0k}=0$), which is supported by most local and cosmological experiments \citep{collaboration2018planck, Abbott_2018}.

The results presented in the paper convince us of two things: First, it is indeed possible to probe dark energy in using GW standard sirens at eLISA. Secondly, it is wiser and more effective to probe null diagnostics instead of the standard EoS parameters for any possible signature of dark energy or the interaction between the dark sectors in eLISA. We would like to reiterate that
the present work deals with almost all types of dark energy models with a wide class of fiducial values chosen from the constraints coming out of existing observational data. Thus, the analysis is robust and the conclusions are more or less generic.

Our paper is organized as follows: In Section 2 we discuss non-interacting dark energy with a comparative study between standard and {\it Om} parametrization in eLISA employing Fisher matrix analysis by taking two widely accepted EoS: namely, the constant $w_0$CDM and the evolving CPLCDM. We extend our analysis for interacting dark sectors in 
Section 3  by redefining the {\it Om} parameter and forecasting on the errors as expected from the two formalisms followed by a comparison between them. In Section 4 we summarise our results and discuss possible open issues. 

\section{Non-Interacting Dark Energy}

\subsection{EoS versus {\it{Om}}  Parametrizations}
The gravitational waveform of supermassive black hole binary (SMBHB) mergers gives GW luminosity distance, $D_L$. Redshift ($z$) has to be obtained from electromagnetic observations. Thus GWs from several SMBHB mergers along with their electromagnetic counterpart give $D_L$ as a function of $z$. Within GR, the luminosity distance of GW is given by, ($c$ is the velocity of light in vacuum)
\begin{equation}\label{D_L}
D_L = \frac{cx}{H_0} \int _1 ^x \frac{dx^{\prime}}{E(x^{\prime})}
\end{equation}
With $x=1+z$, $H_0$ is the value of the Hubble parameter today and in flat universe with CDM, $E(x)$ is approximated as,
\begin{equation}\label{E}
E(x) = \sqrt{\Omega_{0m} x^3 + (1-\Omega_{0m}) x^{3(1+w)}}
\end{equation}

The dark energy EoS parameter $w$ is estimated from the data of $D_L$ vs. $z$ as,
\begin{equation}\label{wEoS}
w = \frac{\frac{2x}{3} \frac{d ln(H_0 E(x))}{dx}-1}{1- \frac{\Omega_{0m}x^3}{[E(x)]^2}}
\end{equation}
For a $\Lambda$CDM universe, $w$ is $-1$. Greater values than $-1$ implies quintessence and $w<-1$ implies phantom.
As is well-known, in the standard dark energy parametrization, this EoS is parametrized and confronted with the luminosity distance data.

In the present paper, we will consider and analyze two distinct representative scenarios that collectively take into account the majority of non-interacting dark energy models available in the literature: 
\begin{enumerate}
\item constant dark energy EoS, parametrized in terms of $w_0$. 

\item dynamical dark energy models represented by the well-known Chevallier-Polarski-Linder (CPL) \citep{Chevallier_2001, PhysRevLett.90.091301} parametrization, where EoS 
is parametrized as,
\begin{center}
$w(z) = w_0 + w_a\left(\frac{z}{1+z}\right)$
\end{center}
\end{enumerate}
In both cases, dark matter is assumed to be cold and non-interacting (CDM).

On the other hand, another recently developed way of parametrizing  dark energy  is via a null diagnostic called the $\textit{Om}$ parameter which  is defined as \citep{OmSahni, Shafieloo_2012}, 
\begin{equation}\label{Om}
\textit{Om}(x) = \frac{[E(x)]^2-1}{x^3-1}
\end{equation}
in flat universe with CDM. {Thus $E(x)$ in this new parametrization technique reduces to} $$E(x)=\sqrt{\textit{Om}(x)x^3+\{1-Om(x)\}}$$
{Thus,} if $\textit{Om}(x)$ is a constant, which means its value is independent of $x$, then the Universe is $\Lambda$CDM and this constant is  $\Omega_{0m}$. {Comparing the above equation with equation \eqref{E} we get $$\textit{Om}(x)=\Omega_{0m}+(1-\Omega_{0m})\frac{x^{3(1+w(x))}-1}{x^3 -1}$$Thus, if $w>-1$,} $\textit{Om}(x)>\Omega_{0m}$  for any $x$ implies quintessence. {Similarly $\textit{Om}(x)>\Omega_{0m}$  implies} a phantom universe. In fact to know the nature of dark energy its not even necessary to know the true value of $\Omega_{0m}$. {This essentially makes \textit{Om} a null diagnostic.} A relation (greater, equal or lesser) between $\textit{Om}(x_1)$ and $\textit{Om}(x_2)$ where $x_1 \neq x_2$ is sufficient to comment on the nature of an universe \citep{OmSahni}. For example if $x_1 > x_2$ then,
\begin{eqnarray*}
    \textit{Om}(x_1)&>&\textit{Om}(x_2) \rightarrow \text{phantom}\\
    \textit{Om}(x_1)&<&\textit{Om}(x_2) \rightarrow \text{quintessence}\\
    \textit{Om}(x_1)&=&\textit{Om}(x_2) \rightarrow \text{$\Lambda$CDM}
\end{eqnarray*}   
The deviation from $\Lambda$CDM is better probed through this technique due to two reasons,
\begin{enumerate}
\item The value of the state parameter $w$ is affected by the error of $\Omega_{0m}$ and the test of the departure from $\Lambda$CDM is sensitive to that error. In this scenario, $\textit{Om}$ offers a null test independent of $\Omega_{0m}$.
\item {In standard parametrization, a comment on the nature of dark energy can be made only after estimation of the EoS parameter.} But, the EoS parameter, $w$, is a function of the second derivative of $D_L$ with respect to $z$, whereas $\textit{Om}$ involves only its first derivative. As a result {studying the  Universe using}  $\textit{Om}$ diagnostic {at two points} is much less {error prone} than the standard method.
\end{enumerate}

Another advantage of using $\textit{Om}$  parametrization is that it does not {explicitly} contain  $H_0$. { However, in order to find $\textit{Om}$ from the observation of luminosity distance, one needs to use \eqref{D_L} and hence the error in $H_0$ (the maximum value of $\Delta H_0/H_0 \sim 0.03$) affects the observations of $\textit{Om}$. Although $\textit{Om}$ is never aimed at estimating $H_0$ from data, the error in $H_0$ introduces an error in $\textit{Om}$, while determining it from the actual observation of GW luminosity distance \citep{OmSahni}.}

In this method the dark energy  is probed by a new parameter $R$, that is constructed as a function of the $\textit{Om}$ parameter at four different redshift points, as \citep{OmSahni}
\begin{equation}\label{R}
R = \frac{\textit{Om}(x_1)-\textit{Om}(x_2)}{\textit{Om}(x_3)-\textit{Om}(x_4)} = \frac{\frac{x_1^{3(1+w)}-1}{x_1^3-1}-\frac{x_2^{3(1+w)}-1}{x_2^3-1}}{\frac{x_3^{3(1+w)}-1}{x_3^3-1}-\frac{x_4^{3(1+w)}-1}{x_4^3-1}}
\end{equation}
with $x_1<x_2$ and $x_3<x_4$. An important point to note here is that like $\textit{Om}$, $R$ is also independent of the present matter density and is ill-defined when the EoS is exactly $-1$  (e.g., for $\Lambda$CDM). 

In principle, to probe dark energy by null diagnostics, one needs to probe $R$ directly from data. However, as already argued, in this article our primary target is to compare the two diagnostics as may be expected from eLISA. So, we would recast $R$ in terms of the  parameters chosen for two representative cases, namely, (i) constant \{$w_0$ \} and (ii) CPL  
\{$w_0, w_a$\}, both with CDM, so that we can make a comparative analysis of the errors between the two diagnostics for the same set of parameters. For these two classes of models, we investigate the prospects of `{\textit{Om}}' over standard formulation in eLISA.

\subsection{Methodology}\label{nimethod}
We are going to use the simplified Fisher matrix analysis to forecast the behavior of `{\textit{Om}}' in eLISA. The Fisher matrix, F for the observations of GW luminosity distance ($D_L$) and redshift ($z$) is defined as \citep{MC}, 
\begin{equation}\label{Fisher Matrix}
    F_{ij} = \sum_{n=\{z\}} \frac{1}{\sigma_n^2} \frac{\partial D_L (z_n)}{\partial \theta_i} \frac{\partial D_L (z_n)}{\partial \theta_j}
\end{equation}
Here $F_{ij}$ is the $ij^{th}$ element of the Fisher matrix and $\{\theta_i\}$ is the set of parameters whose error will be determined in the context of eLISA. $\sigma_n$ is the error in observations of $D_L$ vs. $z$, and $n = \{z\}$ is the distribution of $z$ which gives the redshift points at which the Fisher matrix needs to be evaluated. The inverse of $F$ gives the covariance matrix and the square root of diagonal element of the covariance matrix, $\sqrt{(F^{-1})_{ii}}$ is the required $1-\sigma$ error of parameter $\theta_i$ \citep{MC}.

For our analysis, we take into account $100$ representative redshift points. This gives roughly $10$ times more error in EoS parameters in the standard parametrization regime than reported in \citet{Tamanini_2016}. However, the advantageous nature of {\it Om} over standard technique is independent of the number of data points because the error will decrease by the same amount in both methods as a result of increasing data points. A six link eLISA configuration (like N2A5M5L6) is expected to detect SMBHB mergers between $z=2$ and $z=6$. We also assume a uniform distribution of redshift in the range $[2,6]$. In a realistic scenario,  redshift distribution may not be uniform as the probabilities of the SMBHB merger in all redshifts are not equal.  However, since our job here is to compare the prospects of two diagnostics, considering uniform distribution simplifies our analysis to a great extent. The number of data points and the distribution of redshift can only affect the value of error. In other words, if {\it Om} is deemed better than standard parametrization using 100 uniformly distributed redshift points, the goodness shall remain intact even if any other distribution of redshift is used.

Our primary intention is to do a comparative analysis between the errors in the two formalisms, namely, EoS and null diagnostics, for two distinct classes of non-interacting dark energy models. Thus our choice in  the set of parameters $\{\theta_i\}$ in the two separate formalism are going to be the following: 

\begin{enumerate}
\item In the standard EoS parametrization, we choose: $\{\theta_i\} = \{\Omega_{0m}, w_0\}$ (for $w_0$CDM), and $\{\theta_i\} = \{\Omega_{0m}, w_0, w_a\}$ (for CPLCDM). We then calculate the elements of Fisher matrix $F_{ij}$ using \eqref{Fisher Matrix} via the 
EoS, $w$, and forecast the error in measuring the corresponding parameters in eLISA. 
As pointed out earlier, that encompasses a majority of non-interacting dark energy models. 

\item In the {\it Om} parameter formalism, $R$ plays the role of the EoS. The elements of the Fisher matrix, $F_{ij}$ \eqref{Fisher Matrix} are evaluated in terms of $R$. However, as already argued, in this article our target is to compare between the two diagnostics as may be expected from eLISA. We therefore need to recast $R$ in terms of the same set of parameters as in EoS diagnostics. This will help us make a real comparison between the two formalisms and examine the advantage of one diagnostic over the other if any. 
Now, since $R$ is independent of $\Omega_{0m}$, it always contains one less parameter. Hence, in null diagnostics, $\{\theta_i\} = \{w_0\}$ ($w_0$CDM) and $\{\theta_i\} = \{w_0, w_a\}$ (CPLCDM). This might be one of the  reasons behind better performance of {\it Om}. 
\end{enumerate}

Further, unlike $E(z)$ which is defined at one redshift ($z$), $R$ is dependent on 4 redshifts. However from four data points we need to construct one unique $R$ or else the data set will become dependent hampering our Fisher Matrix analysis. To prevent overcounting a special order of the redshifts is chosen which is given by, $$z_1 < z_2 < z_3 < z_4.$$
{ We choose all possible combination of $\{z_1, z_2, z_3, z_4\}$ from a set of 100 uniformly distributed redshifts for SMBHB mergers, maintaining the order $z_1 < z_2 < z_3 < z_4$ to calculate R. Using it we} estimate the errors on $w_0$ for $w_0$CDM and $(w_0,w_a)$ for CPLCDM and compare it to the errors obtained from standard parametrization.

To avoid any confusion, we would like to stress on the fact that in recasting $R$ in terms of old parameters, we are not going to lose the advantages of null diagnostics in any way, as $F_{ij}$ is still evaluated via $R$ in the second case. So, once we are convinced about the role of null parametrization in reducing the error in eLISA, one can directly make use of $R$ to possibly probe  the nature of dark energy directly irrespective of its EoS.

However, before proceeding to estimate errors on each parameter, a clear knowledge of the possible sources of error on each data point ($\sigma_n$) is necessary. There are primarily five sources of error that affect our analysis.
\begin{enumerate}
    \item There is an experimental error on luminosity distance ($\Delta D_L$) obtained from parameter estimation of the supermassive black hole binary merger waveform. For a 6-link eLISA configuration, the error on luminosity distance is expected to be 10 $\%$ \citep{Klein_2016} of the luminosity distance.
\begin{figure}
\includegraphics[width=\columnwidth]{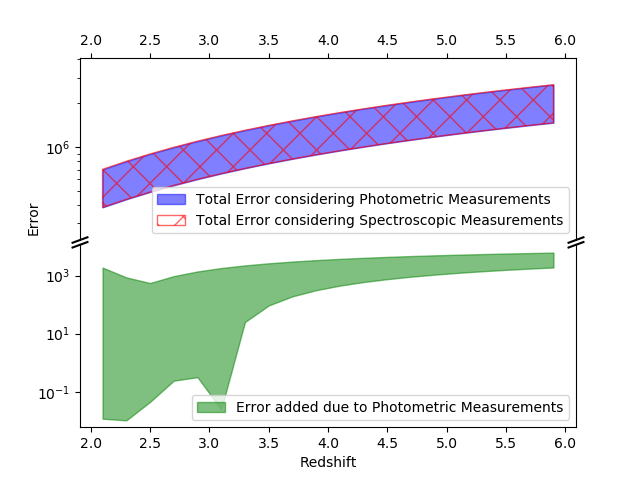}
\caption{ The blue region shows the total error due to spectroscopic measurements and the red crossed region shows the same for photometric measurements. We get a band since all appropriate fiducial values have been considered. Clearly, they overlap. The difference between these two bands is shown by the green region.}
\label{Photometric}
\end{figure}  
    \item The uncertainty on redshift is negligible if spectroscopic techniques are used \citep{Tamanini_2016}. However for several galaxies photometric measurements are easier which introduces an error of the form $0.03 ~\frac{dD_L}{dz} ~(1 +z_n)$ \citep{Dahlen_2013}. {The error introduced due to photometric redshift is at least two orders of magnitude less than the error from other sources and thus contributes insignificantly to the total error as shown in Figure \ref{Photometric}.} In our analysis, we have neglected error from redshift.
    \item Weak lensing introduces systematic error in luminosity distance of standard sirens. The worst case error is of the form \citep{Hirata_2010, Bonvin_2006} $\sigma_{lens} = D_L(z) \times 0.066\left(\frac{1-(1 +z)^{-0.25}}{0.25}\right)^{1.8}$.
    \item Peculiar velocity of GW sources also introduces uncertainty which is given by,
    $$\sigma_v(z) = D_L(z)\left[1 +\frac{c(1 +z)}{H(z)D_L(z)}\right] \frac{\sqrt{ <v>^2}}{c}$$ The root mean square velocity is given by $\sqrt{ <v>^2}$ which we take as 500km/s for our analysis \citep{Kocsis_2006}.
    \item From our analysis we require $E(z)$ for standard parametrization and $R(z_1, z_2, z_3, z_4)$ for {\it{Om}} parametrization. This computation involves the use of the Hubble constant ($H_0$) whose error has to be also taken into account for our analysis. As mentioned earlier their exists tension between local and cosmological measurements of $H_0$. Thus for error estimation in each model, we have taken the value and error of $H_0$ consistent with the fiducial values chosen for that model. {Let us keep in mind that in order to use this parametrization a prior value of $H_0$ is required to calculate $E(z)$ from $D_L(z)$. The importance of $E(z)$ in $\textit{Om}$ parameterization is because $R(z_1,z_2,z_3,z_4)$ requires the value of 4 $E(z)$s. Hence, determining $R(z_1,z_2,z_3,z_4)$ is crucial to obtain equation of state parameters in $\textit{Om}$ parameterization. 
    }
\end{enumerate}

Note also that while using $\textit{Om}$ parametrization we require 4 redshift points. Thus in this case we add the errors for the four points in quadrature. Using this methodology we calculate the error on each parameter in $w_0$CDM and CPLCDM.

\subsection{Results and Analysis}
\subsubsection{$w_0$CDM}
A Fisher matrix analysis for  constant dark energy EoS parameter is presented. Regarding the fiducial values, we take $\{\Omega_{0m}, H_0 \}$ from the combined analysis of two competing sets of observation- Planck 2015 \citep{Planck_2015} as the representative of cosmological observation and galaxy BAO \citep{Alam_2017}, SNeIa \citep{Betoule_2014}, \citet{Riess_2016} (BSH) as the representative of local measurements. The values of $\{\Omega_{0m}, H_0 \}$ with $\Lambda$CDM prior are $\{0.30,~ (68.5 \pm 0.6)\text{km/s/Mpc}\}$ \citep{Bhattacharyya_2019}.

\begin{figure}
	\includegraphics[width=\columnwidth]{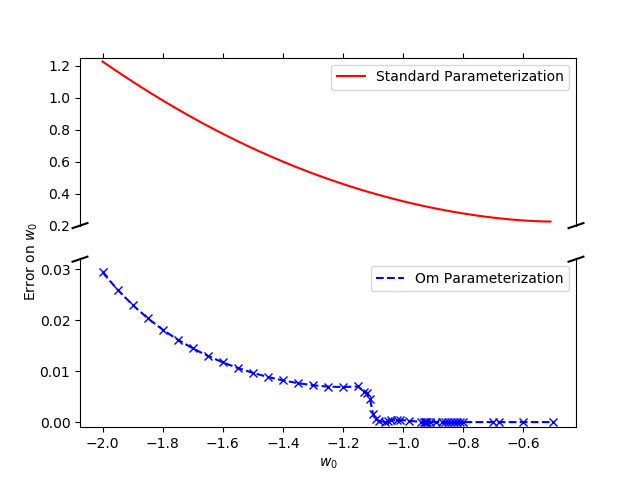}
    \caption{Error on $w_0$ for various fiducial $w_0$s using standard parametrization (upper panel) and the {\it Om} diagnostics (lower panel).{ The advantage of an extended minimum near $w_0=-1$ while using Om diagnostics is that if the Universe has a true value of $w_0$ near -1 (which is supported by most of today's experiments), this parameter has the power to extract $w_0$ with very less error.}}
    \label{fig:w0CDM}
\end{figure}

Figure  \ref{fig:w0CDM} depicts a comparative analysis of the forecasted errors in estimating the parameters between the two diagnostics.
The upper panel represents the error on $w_0$ parameter for standard (EoS) parametrization whereas the lower panel represents the error on the same parameter for null diagnostics. 
From the plot one can readily infer the following:
\begin{itemize}
\item For the same fiducial values, forecasted error using \textit{Om} parameters is about two orders of magnitude less than that of standard parametrization.
\item Error in $w_0$ for phantom fiducial values is greater than non-phantom fiducial values for both parametrizations. This trend is intrinsic to measuring EoS in eLISA and we shall see the same behavior for all other subsequent models. Fortunately enough the greatest error using {\it Om} parameters is an order of magnitude less than the least error using standard parameters for the same number of redshift points. Also, the range of error in $w_0$  using \textit{Om} parameters is about 3\% of the range using standard parameters. 
\item Standard parameters at eLISA shall give the best results if our Universe has a highly quintessence EoS. The advantage of \textit{Om} parameter is that it has a minima in error on $w_0$ for a large range of $w_0$ ($\sim [-1.1,0.5]$). So $\textit{Om}$ parameters give optimal results for a large range of fiducial values of $w_0$.     
\end{itemize}

\subsubsection{CPLCDM}
Likewise, we can do 
a Fisher matrix analysis for models with redshift dependant EoS, which can be described by Chevallier-Polarski-Linder (CPL) parametrization. Like $w_0$CDM, the fiducial values of $\{w_0, w_a, \Omega_{0m}, H_0 \}$ are chosen from the combined analysis of Planck 2015+BSH and from the combined analysis of Planck 2015 \citep{Planck_2015} and \citet{Riess_2016} only (R16). The importance of the last set is that \citet{Riess_2016} is in $4\sigma$ contrast with \citet{Planck_2015}, regarding the value of $H_0$. Hence to understand the deviation from $\Lambda$CDM, this set has a particular importance. The values of $\{w_0, w_a, \Omega_{0m}, H_0 \}$ for Planck 2015+BSH are $\{-1.05, -0.15, 0.29, (69.8 \pm 1.0) \text{km/s/Mpc}\}$ in phantom region and $\{-0.97, 0.04, 0.30, (67.8 \pm 0.7)\text{km/s/Mpc} \}$ in non-phantom region and for Planck 2015+R16 are $\{-1.1, -0.27, 0.26, (74.0 \pm 1.7)\text{km/s/Mpc} \}$ in phantom region and $\{-0.97, 0.03, 0.29, 68.6 ^{+1.3}_{-1.1}\text{km/s/Mpc} \}$ in non phantom region \citep{Bhattacharyya_2019}.

\begin{figure*}
\centering
\includegraphics[scale=0.45,clip,angle=0,keepaspectratio]{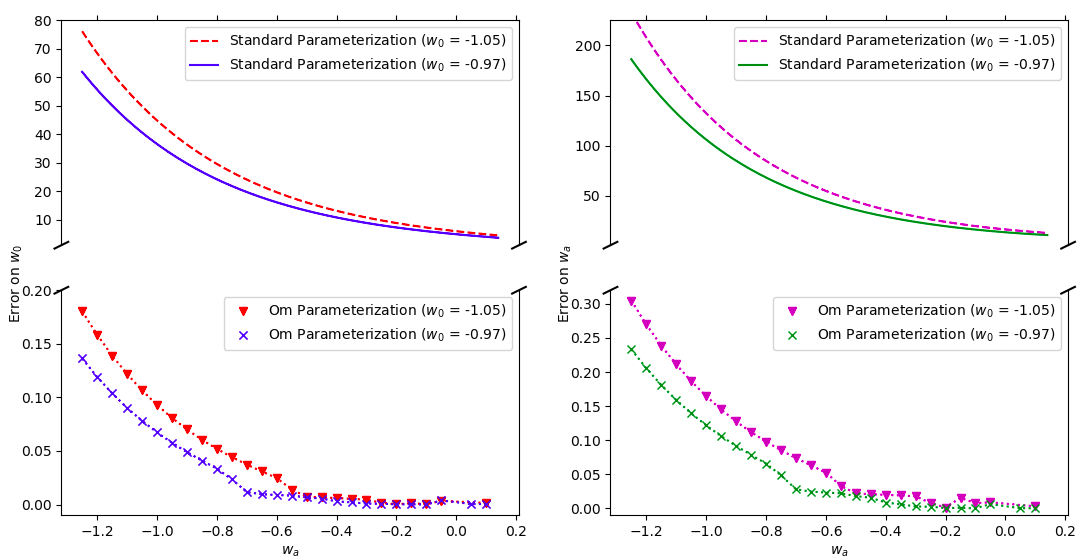}
\caption{Error on $w_0$ (left) and error on $w_a$ (right) vs various fiducial values of $w_a$ is plotted using standard parametrization (top) and the \textit{Om} parametrization (bottom). Two values of $w_0$ are taken: 1) $w_0=-1.05$ (phantom) represented by red (for error on $w_0$)/ magenta (for error on $w_a$). The fiducial values of $H_0$ and $\Omega_{0m}$ are taken as $(69.8 \pm 1.0)$km/s/Mpc and $0.29$ respectively. 2) $w_0=-0.97$ (quintessence) represented by blue (for error on $w_0$)/ green (for error on $w_a$). The fiducial values of $H_0$ and $\Omega_{0m}$ are taken as $(67.8 \pm 0.7)$km/s/Mpc and $0.30$ respectively. Such a choice of fiducial values are made keeping in mind the best fit value of parameters obtained by fitting standard CPLCDM model to Planck + BSH data. The values are taken from \citep{Bhattacharyya_2019}.}
\label{fig:CPLw0}
\end{figure*}

\begin{figure*}
\centering
\includegraphics[scale=0.45,clip,angle=0,keepaspectratio]{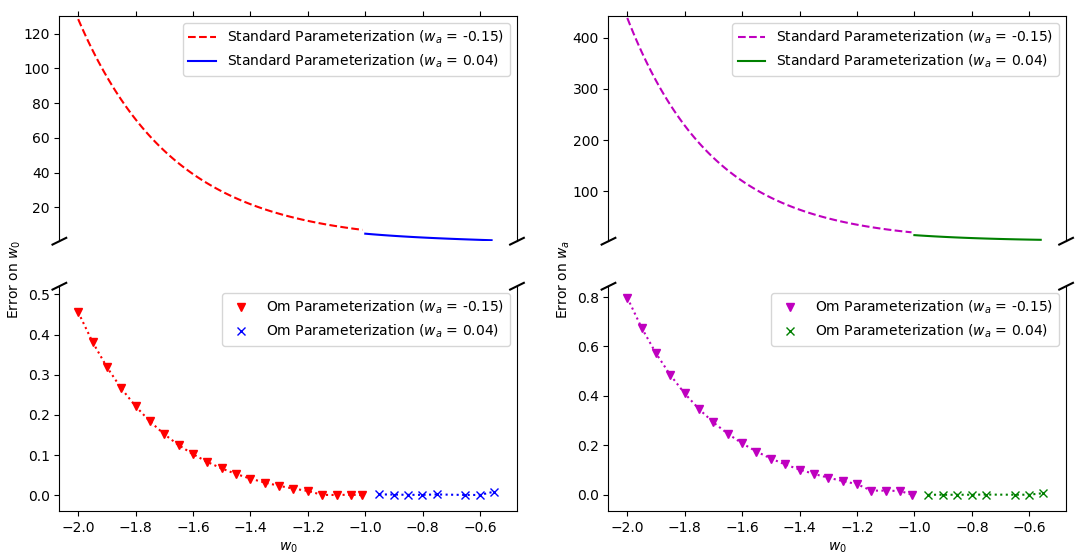}
\caption{Error on $w_0$ (left) and error on $w_a$ (right) vs various fiducial values of $w_0$ is plotted using standard parametrization (top) and the \textit{Om} parametrization (bottom). Two values of $w_a$ are taken: 1) For the phantom region $w_a=-0.15$ represented by red (for error on $w_0$)/ magenta (for error on $w_a$). The fiducial values of $H_0$ and $\Omega_{0m}$ are taken as $(69.8 \pm 1.0)$km/s/Mpc and $0.29$ respectively. 2) For the non-phantom region $w_a=0.04$  represented by blue (for error on $w_0$)/ green (for error on $w_a$). The fiducial values of $H_0$ and $\Omega_{0m}$ are taken as $(67.8 \pm 0.7)$km/s/Mpc and $0.30$ respectively. Such a choice of fiducial values are made keeping in mind the best fit value of parameters obtained by fitting standard CPLCDM model to Planck + BSH data. The values are taken from \citep{Bhattacharyya_2019}. Since the set of fiducial values are different for the phantom and non-phantom region there exists a discontinuity at $w_0 = -1$.}
\label{fig:CPLwa}
\end{figure*}

\begin{figure*}
\centering
\includegraphics[scale=0.45,clip,angle=0,keepaspectratio]{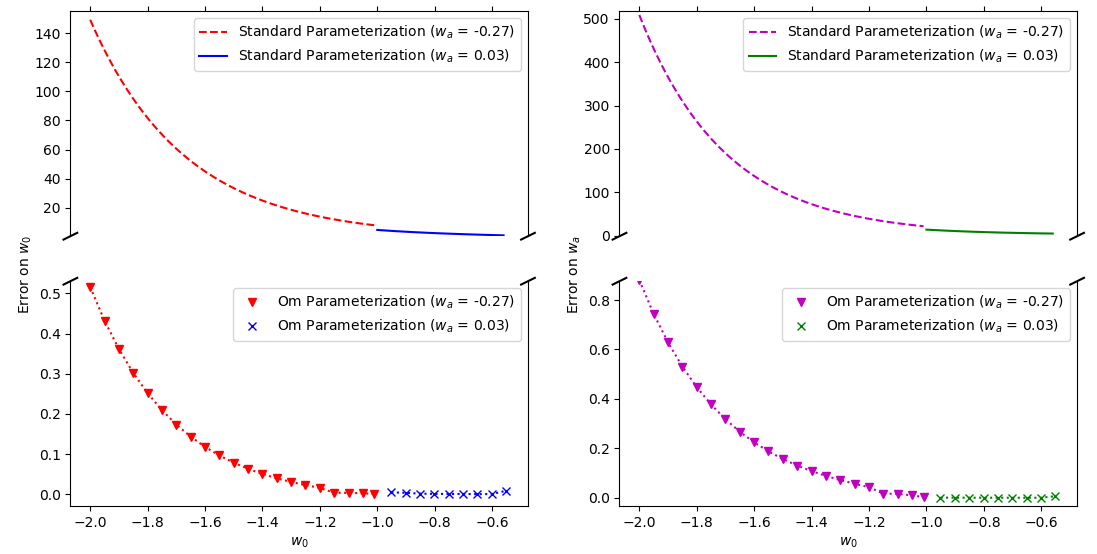}
\caption{Error on $w_0$ (left) and error on $w_a$ (right) vs various fiducial values of $w_0$ is plotted using standard parametrization (top) and the \textit{Om} parametrization (bottom). Two values of $w_a$ are taken: 1) For the phantom region $w_a=-0.27$ represented by red (for error on $w_0$)/ magenta (for error on $w_a$). The fiducial values of $H_0$ and $\Omega_{0m}$ are taken as $(74.0 \pm 1.7)$km/s/Mpc and $0.26$ respectively. 2) For the non-phantom region $w_a=0.03$  represented by blue (for error on $w_0$)/ green (for error on $w_a$). The fiducial values of $H_0$ and $\Omega_{0m}$ are taken as $(68.6 \pm 1.3)$km/s/Mpc and $0.29$ respectively. Such a choice of fiducial values are made keeping in mind the best fit value of parameters obtained by fitting standard CPLCDM model to Planck + R16 data. (Planck + R16 for non-phantom EoS gives $H_0=68.6^{+1.3}_{-1.1}$. However as mentioned previously, we have taken the absolute error on $H_0$ to be $1.3$ to estimate the maximum error on parameters.) The values are taken from \citep{Bhattacharyya_2019}. Since the set of fiducial values are different for the phantom and non-phantom region there exists a discontinuity at $w_0 = -1$. Since Plank and R16 were the first two datasets which did not agree with each other assuming a $\Lambda$CDM model, we have studied the combination of these two separately.}
\label{fig:R16CPLw0}
\end{figure*}

The results have been summarized in figures \ref{fig:CPLw0}  -  \ref{fig:R16CPLw0}.  In all three figures, the upper panels represent the errors on of parameters for EoS parametrization whereas the lower panels represent the same for the same set of parameters for null diagnostics. Taken together, the three figures point at several interesting trends. They are as follows:
\begin{itemize}
    \item Like $w_0$CDM, errors in the phantom region are greater than the non-phantom region for both $w_0$ and $w_a$ using both the methods.
    \item Also for both parameters $w_0$ and $w_a$, errors using \textit{Om} parametrization are significantly less than that of standard parametrization for the same fiducial values. Also, as in the case of $w_0$CDM, variation in the values of errors for \textit{Om} parametrization is less than that of standard parametrization in all the three cases.
    \item In Figure \ref{fig:CPLw0}  the error on $w_0$ and $w_a$ in standard parametrization decreases monotonically for increase in values of $w_a$. Although the trend is similar for \textit{Om}, the curves are more flatter in the range $w_a=[-0.7,0.1]$. This signifies if the true value of $w_a$ is in the range $[-0.7,0.1]$ (which is supported by most experiments till date \citep{PhysRevD.98.083501}), the \textit{Om} parametrization would perform exceedingly well.
    \item Figures \ref{fig:CPLwa}   and \ref{fig:R16CPLw0} also imply that for standard parametrization the error on $w_0$ and $w_a$ decreases monotonically for increase in values of $w_0$. For \textit{Om} parametrization the error curve is much flatter especially in the non-phantom region.
    \item In standard parametrization error on $w_a$ is about 400 $\%$ greater than the error on $w_0$. The biggest advantage of using \textit{Om} parametrization is that the errors on $w_0$ and $w_a$ are of almost the same order.
\end{itemize}

To summarize the above analysis, for 
non-interacting dark energy models, with both constant EoS and redshift dependent EoS, e.g.,  the ones with CPL parametrization, the null diagnostic has better capability to probe dark energy with much less error than the standard EoS formalism, using eLISA.


\section{Interacting Dark Sectors}
Having convinced ourselves on the prospects of null diagnostics in probing dark energy with  much less error than EoS parametrization for 
non-interacting dark energy,  
in this section, we will try to explore the potential of the same for interacting dark sectors.
To accomplish this, we will consider the interaction between two dark fluids, namely, dark matter and dark energy, at the background level. We use a four-parameter phenomenological parametrization that encompasses a wide range of models. At the background level ignoring contributions from baryonic matter, radiation and curvature the evolution equation can be expressed as \citep{Wang_2016},
\begin{eqnarray}
\rho^{\prime}_{\text{dm}} + 3 \mathcal{H} (1+w_{\text{dm}})\rho_{\text{dm}} &=& -aQ\label{Interqction1}\\
\rho^{\prime}_{\text{de}} + 3 \mathcal{H} (1+w_{\text{de}})\rho_{\text{de}} &=& aQ \label{Interqction2}
\end{eqnarray}
where, $\mathcal{H}=\frac{a^\prime}{a}$.
The prime ($\prime$) denotes derivatives with respect to conformal time and $Q$ denotes the coupling or energy transfer between dark matter ($\rho_{\text{dm}}$) and dark energy ($\rho_{\text{de}}$) density. Several phenomenological forms of $Q$ exist in the literature, e.g.,  $Q=-\Gamma \rho_{\text{dm}}$
\citep{B_hmer_2008}, or $Q=H
(\alpha_{\text{dm}}\rho_{\text{dm}}+\alpha_{\text{de}}\rho_{\text{de}})$ \citep{Zimdahl_2001}. Using a particular form for interaction, these type of interacting dark energy (IDE) models have been investigated  to some extent for standard EoS parametrization in the context of eLISA \citep{Caprini_2016}.

However,  we do not aim to choose a particular form for the interaction term $Q$ and hence any particular IDE model for investigating its prospects in eLISA. 
We would rather attempt to put constraints on the interaction between the dark fluids and hence probe interacting dark sectors, in general, using two diagnostics under consideration.  Moreover, using the present data we know that $w_{\text{dm}}$ has to be very small, i.e., it should behave pretty close to CDM. Thus we recast the equations \eqref{Interqction1} and \eqref{Interqction2} in the following form \citep{PhysRevD.78.023505}, 
\begin{eqnarray}
\rho^{\prime}_{\text{dm,eff}} + 3 \mathcal{H} (1+w_{\text{dm,eff}})\rho_{\text{dm,eff}} &=& 0 \label{eq:dmeff}\\
\rho^{\prime}_{\text{de,eff}} + 3 \mathcal{H} (1+w_{\text{de,eff}})\rho_{\text{de,eff}} &=& 0 
\label{eq:deeff}
\end{eqnarray}
where the {\it effective} EoS for the dark components are given by
\begin{eqnarray}
w_{\text{dm,eff}} &=& w_{\text{dm}} + \frac{aQ}{3\mathcal{H} \rho_{\text{dm}}} \\
w_{\text{de,eff}} &=& w_{\text{de}} -\frac{aQ}{3\mathcal{H} \rho_{\text{de}}}
\end{eqnarray}
As demonstrated in a couple of earlier works \citep{Bhattacharyya_2019, PhysRevD.78.023505}, recasting the EoS for the dark components in the above  form will essentially help us to bypass the explicit dependence of the phenomenological interaction term
on observational data, thereby avoiding the necessity of introducing another, model-dependant, parameter in the analysis. 
Consequently, this 
will help us compare the two frameworks (EoS and null diagnostics) in forecast analysis for eLISA from a much wider platform. 

One more advantage of recasting the equations in terms of effective parameters is the following. Even though we start with a set of equations for IDE, these new sets of equations \eqref{eq:dmeff} - \eqref{eq:deeff} can represent a wide class of dark energy models for different values of
the effective parameters. For example,
\begin{itemize}
    \item $w_{\text{dm,eff}}$ = 0; $w_{\text{de,eff}}$ = -1 $\rightarrow$ $\Lambda$CDM
    \item $w_{\text{dm,eff}}$ = 0; $w_{\text{de,eff}}$ = const.($\neq f(z)$) (if greater than -1 then quintessence or less is phantom) $\rightarrow$ cold dark matter with constant dark energy EoS ($w_0$CDM).
    \item $w_{\text{dm,eff}}$ = 0; $w_{\text{de,eff}}$ = $f(z)$ $\rightarrow$ cold dark matter with evolving dark energy EoS ($w_z$CDM,  i.e., CPLCDM).
    \item $w_{\text{dm,eff}}$ $\neq$ 0; $w_{\text{de,eff}}$ = $f(z)$ $\rightarrow$ dark matter with dark energy dark matter interaction or warm dark matter. 
\end{itemize}
Thus the set of equations \eqref{eq:dmeff} - \eqref{eq:deeff} are quite general, and most of the non-interacting and interacting models allowed by present cosmological data, can be constrained by this generalized framework. As a result, they help us to constrain these parameters of different dark energy models accurately and conveniently.

In this scenario, \citet{Caprini_2016} pointed out that only two types of interacting dark energy models can be detected using eLISA. However, we take a parametrized interacting dark energy model. This is due to two causes. Firstly, our parametrization accounts for warm dark matter also, the knowledge of which is particularly important to know about the large scale structure of the Universe. Secondly, our parametrized form is the best possible generalized case which includes most of the models and it is easier to compare other observations with eLISA in this generalized set-up.

In this paper using our simplified three-parameter model, we investigate errors on these parameters using standard parametrization and the {\it Om} parametrization. Thus we define $E(x)$ for interacting dark energy models as,
\begin{equation}\label{EI}
E(x) = \sqrt{\Omega_{0m} x^{3(1+w_{\text{dm}})}+(1-\Omega_{0m}) x^{3(1+w_0+\frac{w_a(x-1)}{x})}}
\end{equation}
As argued, this is the most general form of $E(x)$ that takes into account nearly all types of dark energy models.

In principle both $w_{\text{dm,eff}}$ and $w_{\text{de,eff}}$ can be functions of $z$ and be characterised by CPL like parameters. However we only use CPL like parameters for $w_{\text{de,eff}}$ and treat $w_{\text{dm,eff}}$ as constant. This is because of two reasons. Firstly, all data existing till date allow only a tiny value for $w_{\text{dm,eff}}$, and its variation with redshift is thus also minuscule, thereby making it practically impossible for eLISA to detect it. Secondly opening up both $w_{\text{de,eff}}$ and $w_{\text{dm,eff}}$ would make the parameter space degenerate. Strictly speaking $w_{\text{dm,eff}}$ and $w_{\text{de,eff}}$ are not completely independent for all IDE models under consideration. For example a non-cold dark matter model with no interaction with dark energy may have a non-zero $w_{\text{dm,eff}}$ and $w_{\text{de,eff}}$. However from theoretical perspectives it is not feasible to predict an exact form of interaction (or Hubble Law) and at best we can put some constraints on various parameters from observational data. Here we simply forecast the errors on such parameters at eLISA using standard parametrization and \textit{Om} parametrization.


\subsection{Redefining Null Parameters and Methodology}

As obvious, any non-zero EoS for dark matter and/or dark energy should reflect on the definitions of $\textit{Om}$ and $R$ parameters in the null diagnostics via equation \eqref{EI}. Therefore, in order to accommodate an {\it effective} EoS for dark matter that appears in the set of equations  \eqref{eq:dmeff} - \eqref{eq:deeff}
for the generalized IDE scenario, a minimal modification to the $\textit{Om}$ and $R$ parameters defined for the case non-interacting dark energy is required. Let us denote these new, generalized null parameters by $Om_g$ and $R_g$ respectively. 
\begin{equation}\label{OmI}
Om_g (x) = \frac{[E(x)]^2-1}{x^{3(1+w_{\text{dm}})}-1}
\end{equation}
and consequently,
\begin{equation}\label{RI}
R_g = \frac{Om_g(x_1)-Om_g(x_2)}{Om_g(x_3)-Om_g(x_4)} = \frac{\frac{x_1^{3(1+w)}-1}{x_1^{3(1+w_{\text{dm}})}-1}-\frac{x_2^{3(1+w)}-1}{x_2^{3(1+w_{\text{dm}})}-1}}{\frac{x_3^{3(1+w)}-1}{x_3^{3(1+w_{\text{dm}})}-1}-\frac{x_4^{3(1+w)}-1}{x_4^{3(1+w_{\text{dm}})}-1}}
\end{equation}
with $x_1<x_2<x_3<x_4$ for the same reason discussed earlier.

It is  straightforward  to check that these generalized definitions for null parameters boil down to the corresponding old definitions  \eqref{Om} and \eqref{R}
for the particular case of $w_{dm} =0$. However, these new definitions are  useful for $w_{dm} \neq 0$ as well, be it for warm dark mater or
for an  {\it effective} EoS for dark matter arising from interaction.
So, these generalized null parameters can be used for any analysis involving null diagnostics irrespective of whether or not we are considering CDM. 

Let us recall that in the non-interacting scenario, we did a Fisher matrix analysis for our forecast on eLISA with the set of parameters $\{\theta_i\} $ as defined for standard and null diagnostics as discussed at length in Section \ref{nimethod}.
In the same vein, we employ a similar Fisher matrix analysis for forecasting the errors at eLISA with two major changes $-$ definition of \textit{Om} parametrization is modified and the parameter space has expanded. ($\{\theta_i\} = \{\Omega_{0m}, w_0, w_a,w_{\text{de}}\}$ for  standard parametrization and $\{\theta_i\} = \{w_0, w_a,w_{\text{de}}\}$ for  \textit{Om} parametrization) . The technique for
estimation of errors  for the data points 
and everything else  remains the same as before. 
Further, as before, the fiducial values chosen are the best fit values combining Planck 2015 data with different other observational data, namely, Planck + R16 and 
Planck + BSH for interacting dark sectors \citep{Bhattacharyya_2019}. { The fiducial values are given in Table:\ref{table1} \& Table:\ref{table2}.}

\subsection{Results and Analysis}
\begin{figure*}
\centering
\includegraphics[scale=0.45,clip,angle=0,keepaspectratio]{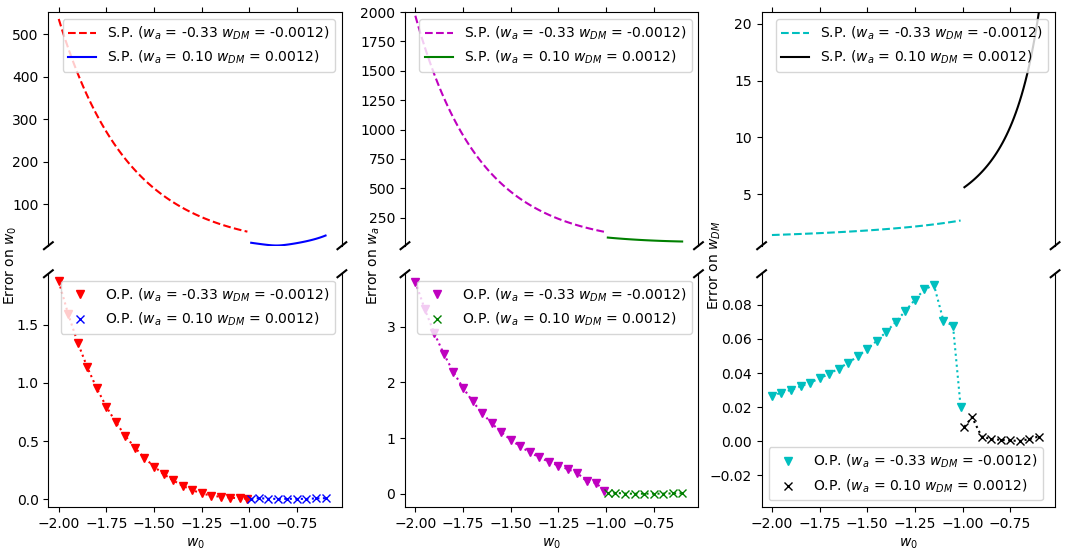}
\caption{Error on $w_0$ (left) and error on $w_a$ (middle) and error on $w_{\text{dm}}$ (right) vs various fiducial values of $w_0$ is plotted using standard parametrization (top) and the \textit{Om} parametrization (bottom). Two pairs of $w_a$,$w_{\text{dm}}$ are taken: 1)For the phantom region: $w_a=-0.33$, $w_{\text{dm}}=-0.0012$  represented by red (for error on $w_0$)/ magenta (for error on $w_a$)/ cyan (for error on $w_{\text{dm}}$). The fiducial values of $H_0$ and $\Omega_{0m}$ are taken as $(69.7 \pm 1.0)$km/s/Mpc and $0.30$ respectively. 2)For quintessence region: $w_a=0.10$, $w_{\text{dm}}=0.0012$ represented by blue (for error on $w_0$)/ green (for error on $w_a$)/ black (for error on $w_{\text{dm}}$). The fiducial values of $H_0$ and $\Omega_{0m}$ are taken as $(68.2 \pm 0.8)$km/s/Mpc and $0.30$ respectively. Such a choice of fiducial values are made keeping in mind the best fit value of parameters obtained by fitting standard CPLCDM model to Planck + BSH data. The values are taken from \citep{Bhattacharyya_2019}. Since the set of fiducial values are different for the phantom and non-phantom region there exists a discontinuity at $w_0 = -1$.}
\label{fig:IDEw0}
\end{figure*}

\begin{figure*}
\centering
\includegraphics[scale=0.45,clip,angle=0,keepaspectratio]{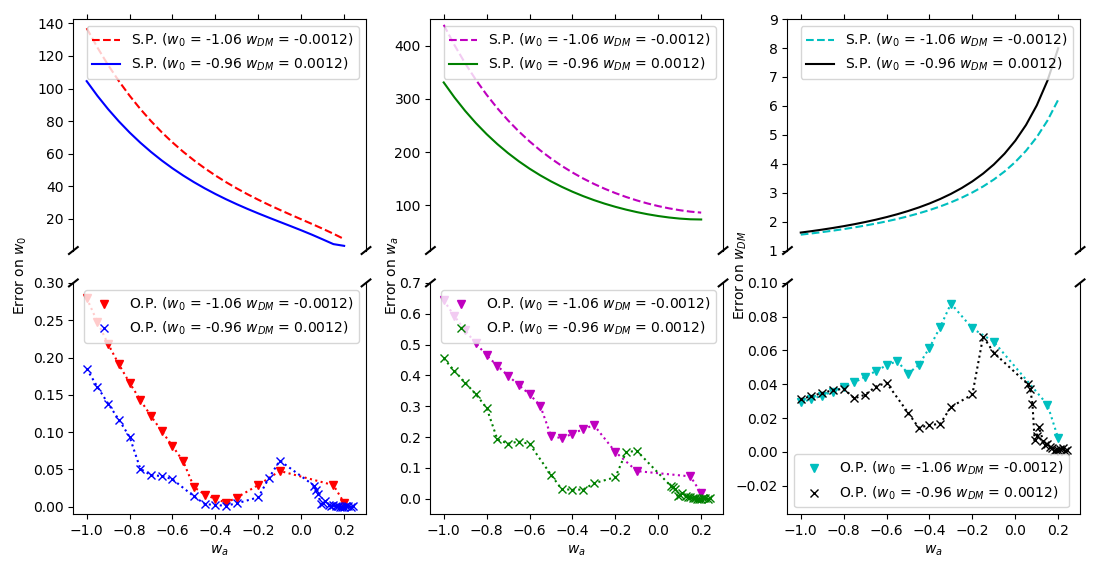}
\caption{Error on $w_0$ (left) and error on $w_a$ (middle) and error on $w_{\text{dm}}$ (right) vs various fiducial values of $w_a$ is plotted using standard parametrization (top) and the \textit{Om} parametrization (bottom). Two pairs of $w_0$,$w_{\text{dm}}$ are taken: 1) $w_0=-1.06$ (phantom), $w_{\text{dm}}=-0.0012$  represented by red (for error on $w_0$)/ magenta (for error on $w_a$)/ cyan (for error on $w_{\text{dm}}$). The fiducial values of $H_0$ and $\Omega_{0m}$ are taken as $(69.7 \pm 1.0)$km/s/Mpc and $0.30$ respectively. 2) $w_0=-0.96$ (quintessence), $w_{\text{dm}}=0.0012$ represented by blue (for error on $w_0$)/ green (for error on $w_a$)/ black (for error on $w_{\text{dm}}$). The fiducial values of $H_0$ and $\Omega_{0m}$ are taken as $(68.2 \pm 0.8)$km/s/Mpc and $0.30$ respectively. Such a choice of fiducial values are made keeping in mind the best fit value of parameters obtained by fitting standard CPLCDM model to Planck + BSH data. The values are taken from \citep{Bhattacharyya_2019}.}
\label{fig:IDEwa}
\end{figure*}

In figures, \ref{fig:IDEw0} and \ref{fig:IDEwa} major results for a Fisher Matrix analysis for an arbitrary constant EoS for dark matter and redshift dependent dark energy EoS are presented.
A comparison between standard EoS parametrization and null diagnostics is clearly visible from the two figures. As in the case of a non-interacting scenario, here also for each plot, the upper panel represents the forecasted errors in measuring the parameters for standard parametrization whereas the lower panel represents the corresponding errors for the same set of parameters for null diagnostics.

Some discussions on 
 figures \ref{fig:IDEw0} and \ref{fig:IDEwa}   are in order. 
\begin{itemize}
    \item As usual, errors using $\textit{Om}$ parametrization is significantly less than standard parametrization for errors on $w_0$ \& $w_a$ . For errors on $w_{\text{dm}}$ efficiency of $\textit{Om}$ parametrization decreases although errors are still less than standard parametrization. This might be because the $\textit{Om}$ parametrization was initially designed to constrain the dark energy EoS.
    However, the generalized definition of $\textit{Om}$, as used here, has the potential to constrain dark matter EoS, if any, as well.
    \item Errors on $w_0$ \& $w_a$ are greater for phantom region using both parametrization. Errors on $w_{\text{dm}}$ vs values of $w_0$/ $w_a$ are greater for quintessence (a nearly monotonic increase from phantom to quintessence) in standard parametrization. However the general trend is somewhat maintained in the case of $\textit{Om}$ parametrization.
    \item Unlike the regular variation of errors on $w_0$, $w_a$ \& $w_{\text{dm}}$ with various fiducial values on $w_a$ using standard parametrization, the variation of errors in $\textit{Om}$ parametrization is erratic with several features. Fortunately the best fit $w_a$ considering Planck+BSH ($w_a=-0.33$ for phantom and 0.10 for quintessence) are situated near the minima of the plots with the well behaved neighbourhood in the lower panel in figure \ref{fig:IDEwa}.
    \item If the present Universe has value of $w_0$ around -2, the $\textit{Om}$ parametrization constrains $w_{\text{dm}}$ very efficiently. This is particularly interesting since Planck combined with R16 hints at $w_0=-2$.
\end{itemize}

\begin{figure*}
\centering
\includegraphics[scale=0.55,clip,angle=0,keepaspectratio]{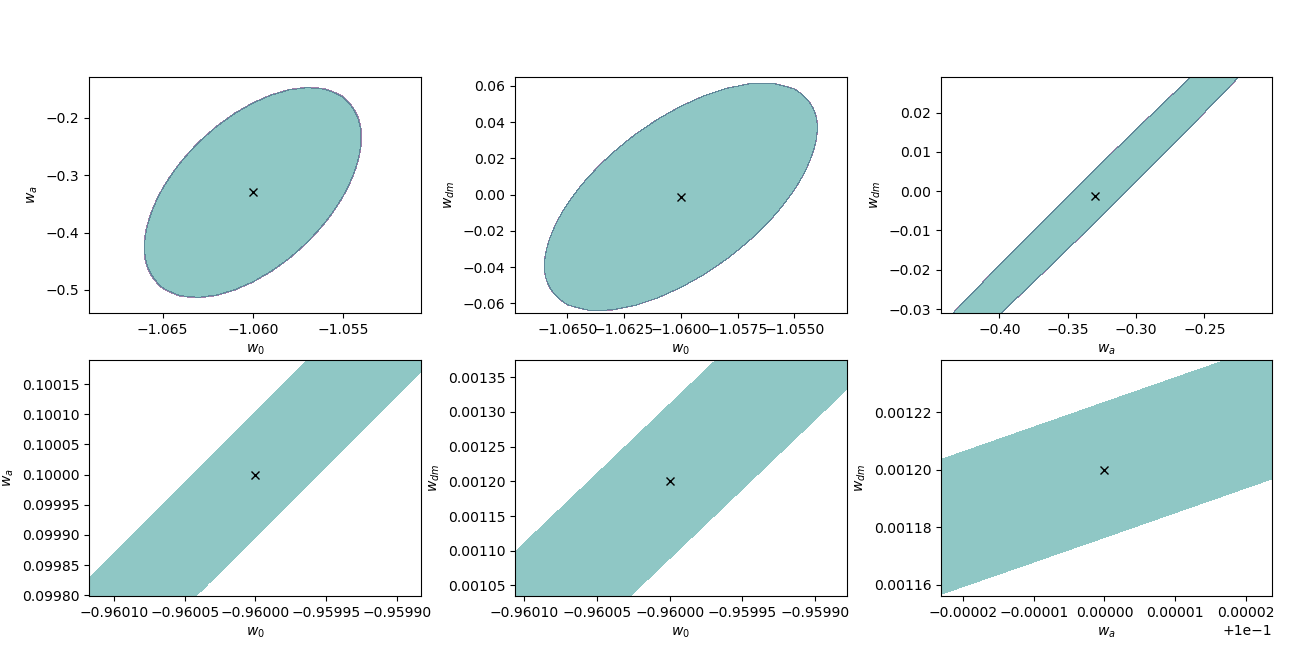}
\caption{Marginalized contour plots for our three-parameter IDE model. The first,second and third column represents 1-$\sigma$ contours for $w_0$-$w_a$, $w_{\text{dm}}$-$w_0$ and $w_{\text{dm}}$-$w_a$ respectively. The contours in the upper row are around the best fit values obtained by fitting our model allowing phantom fiducial values of $w_0$ to Planck+BSH data \citep{Bhattacharyya_2019}. The lower contours are plotted around the best-fit values allowed by a non-phantom $w_0$ in the same dataset. }
\label{fig:Contour}
\end{figure*}

Further, the $1-\sigma$ confidence contour for eLISA for the three parameters for the most general scenario (i.e., interacting dark energy sectors that can in principle take into account almost all the dark energy models in the theoretical framework considered here) have been depicted in figure \ref{fig:Contour}. The figure shows the correlations of the three major parameters under consideration, namely, 
$w_0$, $w_a$ and $w_{\text{dm}}$, separately for phantom and non-phantom fiducial values.

\section{Summary and Outlook}
In this article, we investigated for the prospects of probing dark energy using gravitational waves standard sirens from  SMBHB mergers in the upcoming eLISA mission. The main advantage of such a measurement is that it will make redshifts up to six transparent and thus has the potential to address a couple of unresolved cosmological issues. To this end, we employed Fisher matrix analysis to forecast on the dark energy parameters using two widely accepted formalism, namely, the standard equation of state (EoS) formalism and the model-independent null diagnostics given by {\it Om} parameter. Our main purpose of investigating both formalisms is to do a comparative analysis between them and to find out which one could be more efficient in probing dark energy at high redshift in eLISA by estimating the errors on the same set of parameters separately from the two formalism. We tried with a wide range of fiducial values and different classes of models, namely, the non-interacting $w_0$CDM with constant EoS for dark energy, CPLCDM with evolving EoS, as well as for interacting dark sectors. In the interacting dark sectors, we take a generalized setup used in \citet{Bhattacharyya_2019}, which does not rely on any specific model but represents a class of interacting dark matter-dark energy models via an {\it effective} EoS for dark matter as well as for dark energy. The advantage of doing so is that it can boil down to different (non)interacting models including warm dark matter, for a suitable choice of EoS parameters. 
The fiducial values chosen are the best fit values combining Planck 2015 data with different observational data, namely, Planck + R16 and Planck + BSH \citep{Bhattacharyya_2019}. The present work thus deals with almost all types of dark energy models with a wide class of fiducial values chosen from the constraints coming out of existing observational data. Hence, the analysis is robust.

We have summarized the comparison of forecasted errors between the two diagnostics in tables \ref{table1} and \ref{table2} separately for the phantom and the quintessence regions. The tables show at a glance the forecasted $1-\sigma$ error for eLISA for the two diagnostics for each model and for each combination of datasets under consideration for the corresponding choice of fiducial values.
Together, they are the main results of the present analysis.

\begin{table*}
\caption{1-$\sigma$ error on $w_0$, $w_a$ and $w_{\text{dm}}$ for phantom fiducial values.}
\scalebox{1.3}{
\begin{tabular}{|c|c|c|c|c|c|c|c|c|c|c|}
\hline
\multicolumn{2}{|c|}{} & \multicolumn{3}{c|}{Fiducial Values} & \multicolumn{3}{c|}{Error using Stan. Param.} & \multicolumn{3}{c|}{Error using \textit{Om} Param.} \\ \hline
Data & Model & $w_0$ & $w_a$ & $w_{\text{dm}}$ & $\Delta w_0$ & $\Delta w_a$ & $\Delta w_{\text{dm}}$ & $\Delta w_0$ & $\Delta w_{a}$ & $\Delta w_{\text{dm}}$ \\ \hline
Planck & $w_0$CDM & -1 & - & - & 0.34 & - & - & 7.5e-4 & - & - \\ 
+ & CPLCDM & -1.1 & -0.27 & - & 11 & 29 & - & 3.5e-3 & 0.016 & - \\ 
R16 & IDE & -2.0 & -0.96 & -0.005 & 1800 & 6800 & 1.2 & 6.2 & 11 & 0.014 \\ \hline
Planck & $w_0$CDM & -1 & - & - & 0.35 & - & - & \multicolumn{1}{l|}{4.8e-4} & - & - \\ 
+ & CPLCDM & -1.05 & -0.15 & - & 7.8 & 22 & - & 6.2e-4 & 0.017 & - \\ 
BSH & IDE & -1.06 & -0.33 & -0.0012 & 41 & 150 & 2.6 & 5.4e-3 & 0.19 & 0.068 \\ \hline
\end{tabular}}
\label{table1}
\end{table*}

\begin{table*}
\caption{1-$\sigma$ error on $w_0$, $w_a$ and $w_{\text{dm}}$ for non-phantom fiducial values.}
\scalebox{1.3}{
\begin{tabular}{|c|c|c|c|c|c|c|c|c|c|c|}
\hline
\multicolumn{2}{|c|}{} & \multicolumn{3}{c|}{Fiducial Values} & \multicolumn{3}{c|}{Error using Stan. Param.} & \multicolumn{3}{c|}{Error using \textit{Om} Param.} \\ \hline
Data & Model & $w_0$ & $w_a$ & $w_{\text{dm}}$ & $\Delta w_0$ & $\Delta w_a$ & $\Delta w_{\text{dm}}$ & $\Delta w_0$ & $\Delta w_{a}$ & $\Delta w_{\text{dm}}$ \\ \hline
Planck & $w_0$CDM & -1 & - & - & 0.34 & - & - & 7.5e-4 & - & - \\ 
+ & CPLCDM & -0.97 & 0.03 & - & 4.5 & 13 & - & 0.0018 & 9.1e-4 & - \\ 
R16 & IDE & -0.92 & 0.05 & 0.004 & 7.0 & 58 & 5.9 & 0.013 & 0.006 & 0.001 \\ \hline
Planck & $w_0$CDM & -1 & - & - & 0.35 & - & - & 4.8e-4 & - & - \\ 
+ & CPLCDM & -0.97 & 0.04 & - & 4.5 & 13 & - & 8.2e-4 & 5.1e-4 & - \\ 
BSH & +IDE & -0.96 & 0.10 & 0.0012 & 7.4 & 75 & 6.0 & 4.6e-3 & 0.010 & 9.3e-3 \\ \hline
\end{tabular}}
\label{table2}
\end{table*}

The major outcome of the present analysis can be summarized as follows:
\begin{itemize}
\item \textit{Om} parametrization gives rise to more than  two orders of magnitude improvement in error for different parameters 
and hence is superior to standard parametrization in all cases for all fiducial values.
\item As in the case of standard parametrization, \textit{Om} performs the best for the least number of parameters in the theory. 
\item For dark energy EoS parameter errors decrease with an increase in fiducial values of $w_0$. This trend is generic to eLISA and is valid for all parametrizations. Thus errors in the phantom region are generally more than errors in the quintessence region. 
\item For errors on dark energy vs $w_0$ the trend is reversed with errors increasing as $w_0$ becomes more and more non-phantom in the standard parametrization case. 
\item Results using fiducial values obtained from Planck + R16 do not deviate much qualitatively or quantitatively from those from Planck + BSH and thus our results are immune to fluctuations in fiducial values of $\{\Omega_{0m},H_0\}$.
\end{itemize}

In the paper, we consider a spatially flat universe, as both local and cosmological experiments agree on this issue \citep{collaboration2018planck, Abbott_2018}. However in the presence of non-zero curvature density ($\Omega_{0k}$), the expression of {\it Om} in principle is modified to  \citep{OmSahni},
\begin{center}
    $\overline{Om}(x)=\frac{[E(x)]^2-1}{x^3-1}=\Omega_{0m}+\Omega_{0\Lambda}(\frac{x^{\alpha}-1}{x^3-1})+\Omega_{0k}(\frac{x^2-1}{x^3-1})$
\end{center}
The fractional change of {\it Om} due to introduction of curvature term is \citep{OmSahni},
\begin{center}
    $\frac{\delta Om}{Om}= \frac{\overline{Om}(x)-Om(x)}{Om(x)}= (\frac{\Omega_{0k}}{\Omega_{0m}})\frac{(x+1)}{(x^2+x+1)}$
\end{center}
in the $\Lambda$CDM fiducial universe. It is maximum at the minimum of the redshift points considered which is $z=2$. Hence non-zero curvature term introduces $0.09\%$ error ($\Omega_{0k}=0.001 \pm 0.002$ by Planck 2018+BAO and $\Omega_{0m}=0.315 \pm 0.0007$ \citep{collaboration2018planck} by Planck 2018 alone with base $\Lambda$CDM) in our result, had we used Planck 2018 data.

Our work thus shows the potential of probing the intermediate redshifts with eLISA and null diagnostics with $\textit{Om}$ parametrization. However, this is just a primary work done using $\textit{Om}$ parametrization at eLISA, and some extensions to our work are inevitable. First, a more detailed Fisher matrix analysis can be done by simulating the redshift points using a proper supermassive binary black hole distribution function and noise spectral density for various configurations of eLISA. This will essentially deal with non-uniform datasets and the analysis would presumably be more realistic. Secondly, a Bayesian analysis may even be done. However, we stress that our major findings do not depend on the distribution of redshift used. Thirdly, the fiducial values are chosen for Planck 2015 + R16/BSH as we are unaware of any joint analysis on the generalized dark sectors set up after Planck 2018. Although we do not expect a significant deviation of results using Planck 2018, as this polarization data would at best alter the fiducial values marginally, and thus will not introduce any major change in the forecasted errors, and there will be at most $0.09\%$ change in the estimated error if we take into account the curvature density from Planck 2018, as pointed out in the last paragraph. However, a joint analysis using Planck 2018 + R18 + DES/Pantheon could be done in the future, at least as an update. This will be a two-step process, first, constraining the generalized {\it effective} parameters using those datasets in the line of \citep{Bhattacharyya_2019}, and then doing a Fisher matrix forecast analysis on the parameters for two diagnostics under consideration. 
We plan to take up some of the analyses in the future.

\section*{Acknowledgments}
P.B. and S.K.R. acknowledge Arindom Das and Monabi Basu of Presidency University for computational support. S.K.R. acknowledges Sourodip Dutta of Presidency University for clarifications regarding some statistical issues. { 
The authors also thank the anonymous referee  for constructive suggestions and comments that helped in considerable improvement of the manuscript.
}
\
\section*{Data Availability}
{ The methods used for generating the data required for all the given plots are explained in the manuscript and maybe regenerated with little effort. However, all data generated by the authors for the analysis  will be shared on request to the corresponding author.
The fiducial values of the parameters have been taken from the paper  \citep{Bhattacharyya_2019}  co-authored by one of the authors of the present paper.
}
\bibliographystyle{mnras}
\bibliography{Om}{}

\end{document}